\documentclass[12pt]{article}

\usepackage{amssymb,amsmath,amsfonts,amsthm,eurosym,geometry,ulem,graphicx,caption,color,setspace,sectsty,comment,footmisc,caption,natbib,pdflscape,subfigure,array,hyperref,booktabs,longtable,diagbox,datetime}

\newtheorem*{theorem}{Theorem}
\newtheorem{definition}{Definition}
\newtheorem{proposition}{Proposition}

\title{Information Design of Dynamic Mechanisms}
\author{Chew Soo Hong\thanks{Southwestern University of Finance and Economics and National University of Singapore} \and Wang Wenqian\thanks{Hong Kong University of Science and Technology (Guangzhou)} \thanks{Correpsonding author: wqwang@ust.hk}}
\newdate{date}{20}{05}{2022}
\date{\displaydate{date}}

\begin{document}

\begin{titlepage}
\maketitle
\onehalfspacing

\begin{abstract}

    Two dynamic game forms are said to be behaviorally equivalent if they share the ``same'' profiles of structurally reduced strategies \citep{Battigalli2020}. In the context of dynamic implementation, behaviorally equivalent game forms are interchangeable under a wide range of solution concepts for the purpose of implementing a social choice function. A gradual mechanism \citep{CW2022}, which designs a procedure of information transmission mediated by a central administrator, enables a formal definition of information flow. We provide a characterization of behavioral equivalence between gradual mechanisms in terms of their informational equivalence---each agent is designed the ``same'' information flow. Information flow also helps in defining an intuitive notion of immediacy for gradual mechanisms which is equivalent to their game structures being minimal. Given that the class of gradual mechanisms serves as a revelation principle for dynamic implementation \citep{Li2017,Akbarpour2020,Mackenzie2020,CW2022}, the class of immediate gradual mechanisms provides a refined revelation principle.


\vspace{0.15in}

\end{abstract}
\setcounter{page}{0}
\thispagestyle{empty}
\end{titlepage}

\pagebreak \newpage
\onehalfspacing

\section{Introduction}

The literature on mechanism design has experienced a revival of dynamic implementation, i.e., implementing social choice functions with (possibly) dynamic game forms, enabling the discussion of obviousness \citep{Li2017, CW2022}, simplicity \citep{Bo2021, Pycia2021}, confidentiality \citep{MZ2022, Haupt2022}, and credibility \citep{Akbarpour2020}.\footnote{In this paper, we consider static game form as a special case of dynamic game form in which every agents make only one decision to be made simultaneously.}

Typically in this literature, there is a mechanism designer who wishes to implement a social choice rule that conditions a social outcome on the realization of agents' private types. To do so, the mechanism designer chooses a dynamic game form,\footnote{A dynamic game form provides a procedure for a collection of agents to make interactive decisions which collectively determine the social outcome. It specifies who are active at which point and the collection of options available to them together with some information about the decisions have been made prior to that point. The dynamic game form yields a dynamic game with the realization of each agents' private information at the initial history \citep{Harsanyi1967}.} such that the social choice rule is implemented in equilibrium under certain solution concept when agents participate as players in the derived dynamic game.\footnote{This definition of implementation is also called partial implementation in the literature whereas full implementation requires the set of equilibria corresponds exactly to a social choice correspondence.} 

Ever since \citeauthor{Kuhn1950}'s \citeyearpar{Kuhn1950} formalization of dynamic games, authors have noticed that there exist equivalence relations among dynamic games preserving their strategic features \citep{Thompson1952, Kohlberg1986,Bonannot1992,Elmes1994,Battigalli2020}. Suppose the mechanism designer finds out that a game form implements the social choice rule, she may expect that an equivalent game form works as well. In this paper, we study behavioral equivalence, originally characterized in \cite{Battigalli2020}, in the context of dynamic implementation, especially among a canonical class of dynamic game forms---the gradual mechanism.

Gradual mechanisms have been proposed in the literature to serve as a revelation principle for dynamic implementation under a range of solution concepts such as Bayes-Nash equilibrium \citep{Akbarpour2020}, obvious dominance equilibrium \citep{Li2017,Mackenzie2020}, and its sure-thing principle generalizations \citep{CW2022}. A gradual mechanism can be considered as a procedure of information transmission mediated by a central administrator. It works as follows:



\begin{quote}
    Starting from the initial history, the administrator privately sends specific messages and forms to each of a specific collection of agents. Each form consists of a list of pairwise disjoint categories of the agent's possible type and the agent can check which category his or her type belongs to.\footnote{In the first form each agent receives, the union of the categories is her whole type space. From the second form onwards, the union is the category that the agent checked on the previous form.} The message sent to each agent contains some information about how other agents have checked their forms previously.\footnote{Each history can be identified with an exact description of how each agent has checked a sequence of forms prior to this history. In the message to an agent, he or she is only informed that one history in his or her information set is reached.} The administrator keeps sending forms and messages and collecting the checked forms until she gathers enough information to determine a public outcome.
\end{quote}


One advantage to work with gradual mechanisms is that information flows can be naturally defined since each history explicitly corresponds to a subset of the space of type profiles. Upon arriving at a certain history, the administrator obtains information about agents' individual types reflected in their latest decisions and composes a message incorporating such information for each active agent in that history.

Based on information flow, we offer a definition of informational equivalence between gradual mechanisms and show that it delivers a characterization of behavioral equivalence among gradual mechanisms. Roughly speaking, informational equivalence requires two gradual mechanisms design the ``same'' information flows for each agents.

Information flow also helps to define a notion of immediacy for gradual mechanisms which corresponds to minimality of game structures as defined in \cite{Battigalli2020}. Therefore, as each gradual mechanism is informationally equivalent to a unique immediate gradual mechanism, the collection of immediate gradual mechanisms then provides a refined revelation principle for dynamic implementation.

\section{Behavioral Equivalence in Dynamic Implementation}

There is a finite group of agents $N$ who have interest on a set of public outcomes $X$. Each agent $i\in N$ has a private type space $\Theta_i$. Conventionally, we write a type profile as $\theta = (\theta_i, \theta_{-i})\in \Theta = \prod_{i\in N} \Theta_i$ where $\theta_{-i}$ is a type profile of agents other than $i$.

The designer wishes to implement a social choice function $f:\Theta\rightarrow X$ that conditions a public outcome on agents' type profile through equilibrium plays by agents in a dynamic game form with perfect recall.


A dynamic game form $G$ with perfect recall can be described by the following tuple \[G = (\overline{H}, \{A_i, \boldsymbol{H}_i\}_{i\in N}, \mathcal{X}).\] We explicitly allow for simultaneous actions and model histories using sequences of action profiles \citep{osborne1994course, Battigalli2020}. The following table lists some standard notations that will be used.\footnote{A formal definition of dynamic game forms is provided in \ref{sec:app_definitions}.}

\begin{footnotesize}
\begin{longtable}{l l l}
    \toprule
    Name & Notation/Definition & Generic Element \\
    \midrule
    Actions of $i\in N$ & $A_i$ & $a_i$ \\
    \midrule
    Non-empty Action Profiles & $\bigcup_{\emptyset\neq M\subset N} \prod_{i\in M} A_i$ & $a = (a_i)_{i\in M}$ \\
    \midrule
    Histories & $\overline{H}\subseteq \bigcup_{t = 0}^\mathcal{T} A^T$ is a tree. & $h = \varnothing$ or $h = (h^{(1)}, \ldots, h^{(T)})$ \\
    \midrule
    Precedence Relation over Histories & $\preceq$ & $\underline{h}\preceq h\preceq \overline{h}$ \\
    \midrule
    Terminal Histories & $\preceq$-maximal histories $Z\subseteq \overline{H}$ &  \\
    \midrule
    Non-terminal Histories & $H = \overline{H}\backslash Z$ &  \\
    \midrule
    Outcome function & $\mathcal{X}:Z\rightarrow X$ &  \\
    \midrule
    Active-player Correspondence & $\mathbb{P}: H\twoheadrightarrow N$ & \\
    \midrule
    Active History of agent $i$ & $H_i = \{h\in H: i\in \mathbb{P}(h)\}$ &  \\
    \midrule
    Information Sets & $\boldsymbol{H}_i$ is a partition of $H_i$ & $\boldsymbol{h}_i$ \\
    \midrule
    Available Actions at $\boldsymbol{h}_i$ or $h\in \boldsymbol{h}_i$ & $A_i(h) = A_i(\boldsymbol{h}_i)\subseteq A_i$ &  \\
    \midrule
    Space of Interim Strategies & $S_i$ & $s_i:H_i\rightarrow A_i$ ($s_i:\boldsymbol{H}_i\rightarrow A_i$) \\
    \bottomrule

\end{longtable}
\end{footnotesize}

We use $\Gamma = (G, \Theta)$ to model an incomplete information game in which each agent $i$ only knows her own type, i.e., when $\theta$ is realized, agent $i$ knows only $\theta_i$. A strategy of agent $i$ in $\Gamma$ is then modeled by $\mathbb{S}_i:\Theta_i\rightarrow S_i$ (with $\mathbb{S} = \prod_{i\in N} \mathbb{S}_i: \Theta\rightarrow S$ being a strategy profile) where $\mathbb{S}_i(\theta_i) = s_i$ is the interim strategy agent $i$ adopts when her private type is $\theta_i$.\footnote{The following analysis can also be extended to mixed strategies, i.e., $\mathbb{S}_i:\Theta_i\rightarrow M_i$ in which $M_i$ is the space of mixed strategies arising from $S_i$.} We say that a social choice function $f$ is implemented by $(G, \Theta, \mathbb{S})$ if for each $\theta\in\Theta$ it is the case that $f(\theta) = \mathcal{X}(\mathbb{S}(\theta))$ with $\mathbb{S}$ being an equilibrium of $(G, \Theta)$ under some solution concept.

Starting from a non-terminal history $h\in H$, after specifying the profile of interim strategies $s\in S = \prod_{i\in N} S_i$, a unique terminal history $Z(h, s)$ along with the associated outcome $\mathcal{X}(h, s)$ will be determined. In these expressions, we will omit the history $h$ if it is the initial history, i.e., $Z(\varnothing, s)$ will be denoted as $Z(s)$, and its associated outcome will simply be denoted as $\mathcal{X}(s)$. Following \cite{Battigalli2020}, we first define behavioral equivalence between two interim strategies.

\begin{definition}
    Two interim strategies $s_i, s_i'\in S_i$ are behaviorally equivalent if $Z(s_i, s_{-i}) = Z(s_i', s_{-i})$ for all $s_{-i}$.
\end{definition}

Define $H(s_i) = \{h\in \overline{H}: \exists s_{-i}\in S_{-i} \mbox{ s.t. } h\preceq Z(s_i, s_{-i})\}$ to be the collection of histories that is consistent with the the interim strategy $s_i$ of agent $i$. Define also $\boldsymbol{H}_i(s_i) = \{\boldsymbol{h}_i\in \boldsymbol{H}_i: \exists h\in H(s_i) \mbox{ s.t. } h\in \boldsymbol{h}_i\}$ as the collection of information sets of agent $i$ consistent with her interim strategy $s_i$. When $s_i$ and $s_i'$ are behaviorally equivalent, we can observe that $H(s_i) = H(s_i')$ and $s_i(\boldsymbol{h}_i) = s_i'(\boldsymbol{h}_i)$ for all $\boldsymbol{h}_i\in \boldsymbol{H}_i(s_i) = \boldsymbol{H}_i(s_i')$.\footnote{Behavioral equivalence between two interim strategies is also known as realization equivalence in the literature.}

Behavioral equivalence between interim strategies delivers an equivalence class on $S_i$ for each agent $i$, denoted by $\mathcal{S}_i$ with generic element $\boldsymbol{s}_i$ called a (structurally) \textit{reduced strategy}. Notice that $\boldsymbol{H}_i(\boldsymbol{s}_i)$, $Z(\boldsymbol{s})$, and $\mathcal{X}(\boldsymbol{s})$ are all well-defined. We can now define behavioral equivalence between two dynamic game forms when they provide the ``same'' collection of structurally reduced strategies for each agent.\footnote{\cite{Battigalli2020} define behavioral equivalence between game structures which are game forms without outcome functions. For our investigation of dynamic mechanisms, we define for dynamic game forms.}

\begin{definition}
    Two dynamic game forms $G$ and $G'$ are behaviorally equivalent if there exist bijection $\sigma_i: \boldsymbol{S}_i\rightarrow \boldsymbol{S}_i'$ for each agent $i$ and bijection $g:Z\rightarrow Z'$ such that $\mathcal{X}(z) = \mathcal{X}'(g(z))$ and $g(Z(\boldsymbol{s})) = Z'(\sigma(\boldsymbol{s}))$ in which $\sigma(\boldsymbol{s}) = (\sigma_i(\boldsymbol{s}_i))_{i\in N}$.
\end{definition}


Many solution concepts used in the dynamic implementation literature, such as Bayes-Nash equilibrium, weak dominance equilibrium, obvious dominance equilibrium and its sure-thing principle generalizations, do not distinguish between behaviorally equivalent strategies. That is to say, when $\mathbb{S}$ is an equilibrium, so is $\mathbb{S}'$ such that $\mathbb{S}'_i(\theta_i)$ and $\mathbb{S}_i(\theta_i)$ are behaviorally equivalent for all agent $i\in N$ and $\theta_i\in\Theta_i$. It can be shown that if a dynamic game form $G$ implements a social choice function $f$ under one of these solution concepts, any dynamic game form $G'$ that is behaviorally equivalent with $G$ also implements $f$ under the same solution concept.

Notice that if the mechanism designer relies on a solution concept that distinguishes between behaviorally equivalent strategies (not invariant to the transformations preserving behavioral equivalence between dynamic game forms), such as perfect/sequential equilibrium \citep{Selten1975, Kreps1982}, it remains possible that behaviorally equivalent game forms are interchangeable for the purpose of implementing a social choice function. Suppose such an equilibrium $\mathbb{S}$ exists in $\Gamma = (G, \Theta)$. In checking the interchangeability of a behaviorally equivalent $G'$, all that is needed is the existence of $\mathbb{S}'$ in $\Gamma' = (G', \Theta)$ such that $\mathbb{S}(\theta)\in \boldsymbol{s}$ implies that $\mathbb{S}'(\theta)\in \sigma(\boldsymbol{s})$ for each $\theta\in\Theta$.


\section{Information Flows in Gradual Mechanisms}

For any $h\in \overline{H}$ of a dyanamic game form, let $h_i$ be the sequence of actions agent $i$ has taken up to $h$, and let $h_i^{(-1)}$ denote the last action taken by agent $i$ in $h$. A gradual mechanism is formally defined below.

\begin{definition}

    A \textbf{gradual mechanism} $G$ for social choice function $f$ is a dynamic game form such that:

    \begin{enumerate}
        \item For each agent $i\in N$, the collection of feasible actions are non-empty subsets of her possible types, i.e., $A_i = 2^{\Theta_i}\backslash \{\varnothing\}$.
        \item For each agent $i\in N$ and any history $h\in H_i$ where agent $i$ is active, the available actions $A_i(h)$ are pairwise disjoint subsets of $\Theta_i$ whose union is the collection of remaining types agent $i$ can indicate to be, i.e., $a_i\cap a_i'=\varnothing$ for any $a_i, a_i'\in A_i(h)$ and $\bigcup A_i(h) = \bigcap h_i$.\footnote{By the first condition, $h_i$ is a sequence of subsets of agent $i$'s private type space whose intersection is denoted as $\bigcap h_i$. Note that $\bigcup A_i(h) = \Theta_i$ if $h_i = \varnothing$, i.e., the whole type space is still possible when no action has been taken. Also note that when $h_i\neq \varnothing$, it is the case that  $\bigcap h_i = h_i^{(-1)}$.}
        \item The public outcome assigned to any terminal history $z\in Z$ is aligned with the social choice function based on information accrued up to $z$, i.e., for any $\theta\in \prod_{i\in N} z_{i}^{(-1)}$, it is the case that $\mathcal{X}(z) = f(\theta)$.
    \end{enumerate}

\end{definition}

The applicability of gradual mechanism as a revelation principle has been discussed under several solution concepts, and in fact any dynamic implementation $(G, \Theta, \mathbb{S})$ induces a gradual mechanism implementation $(G^*, \Theta, \mathbb{T})$ under pruning and relabeling in which $\mathbb{T}$ is a profile of type-revealing strategies \citep{Li2017,Akbarpour2020,Mackenzie2020,CW2022}.

Histories and information sets in any gradual mechanism $G = (\overline{H}, \{A_i, \boldsymbol{H}_i\}_{i\in N}, \mathcal{X})$ are clearly associated with the information being transmitted. At history $h\in \overline{H}$, the information, if any, collected by the administrator about agent $i$'s private type is agent $i$'s last decision $\Theta_i(h) = h_i^{(-1)}$. In case that agent $i$ has not made any decision at history $h$, the administrator collects no information about her, i.e., $\Theta_i(h) = \Theta_i$. The overall information collected by the administrator about agents' type profiles is therefore $\Theta(h) = \prod_{i\in N}\Theta_i(h)$. 

We refer to $\mathbb{F} = \{\Theta(h)\}_{h\in \overline{H}}$ as administrator's \textit{information flow} and offer the following observations. For any $h\in H$, let $\sigma(h)$ be the collection of its immediate successors.

\begin{proposition}
    \label{prop:flow}
    The administrator's information flow $\mathbb{F} = \{\Theta(h)\}_{h\in \overline{H}}$ has the following properties:
    \begin{enumerate}
        \item $\Theta(\varnothing) = \Theta$.
        \item $\{\Theta(h): h\in \sigma(\underline{h})\}$ is a partition of $\Theta(\underline{h})$.
        \item $\underline{h}\preceq h$ if and only if $\Theta(\underline{h})\supseteq \Theta(h)$.
        \item If $h'$ and $h$ are not $\preceq$-comparable, then $\Theta(h')\cap \Theta(h)=\varnothing$.
    \end{enumerate}
\end{proposition}

The first property says that before playing the game, the administrator knows nothing about agents' type profile. The second property is a direct consequence of the non-cooperative or decentralized nature of decision making in games. These two properties together depict the histories in a gradual mechanism as a partition tree of $\Theta$. The third and fourth properties are implications of the first two properties. 

Upon collecting information $\Theta(h)$ at history $h$, the administrator sends messages and new forms to every active agent $i\in \mathbb{P}(h)$ to collect new information about them. Since agent $i$ is informed only that one history in her current information set $\boldsymbol{h}_i$ (i.e., $h\in \boldsymbol{h}_i$) has been reached, the information provided to agent $i$ is $\Theta_{-i}(\boldsymbol{h}_i) = \bigcup_{h\in \boldsymbol{h}_i} \Theta_{-i}(h)$. Arising from perfect recall, for any two histories $h, h'\in \boldsymbol{h}_i$ sharing the same information set, it is the case that $\Theta_i(h) = \Theta_i(h')$, which can be denoted as $\Theta_i(\boldsymbol{h}_i)$. Therefore, $\Theta(\boldsymbol{h}_i) = \bigcup_{h\in \boldsymbol{h}_i} \Theta(h) = \Theta_{-i}(\boldsymbol{h}_i)\times \Theta_i(\boldsymbol{h}_i)$, capturing agent $i$'s information associated with the information set $\boldsymbol{h}_i$, can be divided between the information agent $i$ acquires from administrator's message, i.e., $\Theta_{-i}(\boldsymbol{h}_i)$, and information she provides to the administrator, i.e., $\Theta_i(\boldsymbol{h}_i)$.

To ease exposition of the properties of agents' information flows, we introduce agents' information sets on terminal histories. For each agent $i$, let $\boldsymbol{Z}_i$ be a partition of the terminal histories such that the extended information sets $\overline{\boldsymbol{H}}_i = \boldsymbol{H}_i\cup \boldsymbol{Z}_i$ still satisfy perfect recall. Therefore, $\overline{\boldsymbol{H}}_i(s_i)$ is well defined for each interim strategy $s_i\in S_i$ and $\overline{\boldsymbol{H}}_i(\boldsymbol{s}_i)$ for each reduced strategy $\boldsymbol{s}_i\in\boldsymbol{S}_i$. Information sets on terminal histories could serve as a useful tool in studying confidentiality and credibility. e.g., distinguishing between simultaneous decisions that turn public or remain private.\footnote{They correspond, respectively, to a finest and a coarsest partition of the collection of terminal histories, see \cite{Akbarpour2020}. Notice that since information sets on terminal histories do not have strategic effects, any such extension is compatible with the subsequent analysis.}  

Notice that $\overline{\boldsymbol{H}}_i$ forms an arborescence with $\underline{\boldsymbol{h}}_i\preceq \boldsymbol{h}_i$ defined by existence of $\underline{h}\in \underline{\boldsymbol{h}}_i$ and $h\in \boldsymbol{h}_i$ such that $\underline{h}\preceq h$, i.e., the precedence relation $\preceq$ over $\overline{\boldsymbol{H}}_i$ totally orders the predecessors of each member of it. For any $\underline{\boldsymbol{h}}_i\in\boldsymbol{H}_i$, let $\sigma(\underline{\boldsymbol{h}}_i)$ be the collection of its immediate successors. Suppose $a_i\in A_i(\underline{\boldsymbol{h}}_i)$. Let $\sigma_{a_i}(\underline{\boldsymbol{h}}_i)$ be the sub-collection of its immediate successors $\boldsymbol{h}_i$ after taking action $a_i$, meaning that $\Theta_i(\boldsymbol{h}_i) = a_i$ in gradual mechanisms.

We refer to $\mathbb{F}_i = \{\Theta(\boldsymbol{h}_i): \boldsymbol{h}_i\in \overline{\boldsymbol{H}}_i\}$ as the \textit{information flow} for agent $i$.

\begin{proposition}
    \label{prop:Flows}
    For each agent $i$, the information flow $\mathbb{F}_i$ satisfies the following properties.
    \begin{enumerate}
        \item For any $\underline{\boldsymbol{h}}_i\in \boldsymbol{H}_i$, $\{\Theta(\boldsymbol{h}_i)\}_{\boldsymbol{h}_i\in \sigma(\underline{\boldsymbol{h}}_i)}$ is a partition of $\Theta(\underline{\boldsymbol{h}}_i)$.
        \item For any $\underline{\boldsymbol{h}}_i\in \boldsymbol{H}_i$ and any $a_i\in A_i(\underline{\boldsymbol{h}}_i)$, $\{\Theta_{-i}(\boldsymbol{h}_i)\}_{\boldsymbol{h}_i\in \sigma_{a_i}(\underline{\boldsymbol{h}}_i)}$ is a partition of $\Theta_{-i}(\underline{\boldsymbol{h}}_i)$.
    \end{enumerate}
\end{proposition}

The second property says that after reporting any information back to the administrator, an agent is expected to acquire refined information about other agents' type profile.



\section{Informational Equivalence and Immediacy}

For general game structures, \cite{Battigalli2020} characterize behavioral equivalence with elementary transformations, known as ``Interchanging of Simultaneous Moves'' and ``Coalescing Moves/Sequential Agent Splitting''. More specifically, two game structures are shown to be behaviorally equivalent if they are connected to the same minimal game structure through a sequence of ``coalescing'' and ``simultanizing'' transformations.\footnote{We provide their adaptation to gradual mechanisms in \ref{sec:Elementary_Transformation}.}

The concept of information flow enables an intuitive definition of immediacy for gradual mechanisms, corresponding to minimality of game structures characterized in terms of absence of ``coalescing'' and ``simultanizing'' opportunities.

\begin{definition}

    A gradual mechanism $G$ is \textbf{immediate} in terms of information flow if for each agent $i$ the following two conditions hold:
    
    \begin{enumerate}
        \item For any two $\underline{\boldsymbol{h}}_i, \boldsymbol{h}_i\in \boldsymbol{H}_i$ with $\underline{\boldsymbol{h}}_i\prec \boldsymbol{h}_i$, it is the case that $\Theta_{-i}(\underline{\boldsymbol{h}}_i) \supsetneq \Theta_{-i}(\boldsymbol{h}_i)$;
        \item For any $h\in H$ and $\boldsymbol{h}_i\in \boldsymbol{H}_i$, if $\Theta_{-i}(h) \subseteq \Theta_{-i}(\boldsymbol{h}_i)$ and $\Theta_i(h) = \Theta_{i}(\boldsymbol{h}_i)$, then $h\in \boldsymbol{h}_i$.
    \end{enumerate}
    
\end{definition}

This definition can be interpreted intuitively. In the first condition, each agent $i$ is not asked by the administrator for new information unless she is provided some new information about other agents. The second condition says that the administrator will approach each agent immediately after enough information is accrued to compose a new message for her.

We can further provide a characterization of behavioral equivalence between two gradual mechanisms with the following informational equivalence defined in terms of information flow.

\begin{definition}
    \label{def:b_equivalence}

    Two gradual mechanisms $G$ and $G'$ are \textbf{informationally equivalent} if they implement the same social choice function and:

    \begin{enumerate}
        \item for each $z\in Z$, there exists $z'\in Z'$ such that $\Theta(z) = \Theta(z')$;
        \item for each agent $i$, for each information set $\boldsymbol{h}_i\in \boldsymbol{H}_i$, there exists $\boldsymbol{h}_i'\in \boldsymbol{H}_i'$ such that $\Theta_{-i}(\boldsymbol{h}_i) = \Theta_{-i}(\boldsymbol{h}_i')$ and $\Theta_i(\boldsymbol{h}_i) \subseteq \Theta_i(\boldsymbol{h}_i')$;
        \item for each agent $i$, for each information set $\boldsymbol{h}_i'\in \boldsymbol{H}_i'$, there exists $\boldsymbol{h}_i\in \boldsymbol{H}_i$ such that $\Theta_{-i}(\boldsymbol{h}_i) = \Theta_{-i}(\boldsymbol{h}_i')$ and $\Theta_i(\boldsymbol{h}_i') \subseteq \Theta_i(\boldsymbol{h}_i)$.
    \end{enumerate}
   
\end{definition}

The first condition says that the administrator collects the same information at the end of two informationally equivalent gradual mechanisms. The second and third conditions apply to agents. The information flows in $G$ and $G'$ are the same in the sense that for any information set in one of the two gradual mechanism, there exists another in the other on which the agent acquires the same information while providing less information to the administrator. An implication of these two conditions is that there exists, for each information set $\boldsymbol{h}_i$ in either of two informationally equivalent gradual mechanisms, the same information set $\underline{\boldsymbol{h}}_i$ (in terms of the information acquired and provided by the agent $i$) in both gradual mechanisms with $\Theta_{-i}(\boldsymbol{h}_i) = \Theta_{-i}(\underline{\boldsymbol{h}}_i)$ and $\Theta_{i}(\boldsymbol{h}_i) \subseteq \Theta_{i}(\underline{\boldsymbol{h}}_i)$. Notice that informational equivalence does not directly place requirements on the information accrued at non-terminal histories of a gradual mechanism.

We are now ready to present our main result.

\begin{theorem}
    Any gradual mechanism $G$ is informationally equivalent to a unique immediate gradual mechanism. Furthermore, two gradual mechanisms $G$ and $G'$ are informationally equivalent if and only if they are behaviorally equivalent.
\end{theorem}

\begin{proof}

Given a gradual mechanism $G$, we can construct an immediate gradual mechanism $G^*$ that is informationally equivalent to it using the following \textit{instantizing} procedure. First, for each agent $i$ and each terminal history $z$, find the collection of information sets of agent $i$ on the path to $z$, given by $\overline{\boldsymbol{H}}_{i,z} = \{\boldsymbol{h}_i\in\overline{\boldsymbol{H}}_i: \exists h\in \boldsymbol{h}_i \mbox{ s.t. } h\preceq z\}$. Second, eliminate from each $\overline{\boldsymbol{H}}_{i,z}$ the information sets on which the agent is acquiring the same information but providing more, i.e., derive $\overline{\boldsymbol{H}}_{i,z}^* = \{\underline{\boldsymbol{h}}_i\in \overline{\boldsymbol{H}}_{i,z}: \forall \boldsymbol{h}_i\in \boldsymbol{H}_{i,z}, \Theta_{-i}(\underline{\boldsymbol{h}}_i) = \Theta_{-i}(\boldsymbol{h}_i) \mbox{ implies } \Theta_{i}(\underline{\boldsymbol{h}}_i) \subseteq \Theta_{i}(\boldsymbol{h}_i)\}$. Third, obtain the instantized information flow of each agent on the remaining information sets $\mathbb{F}_i^* = \{\Theta(\boldsymbol{h}_i): \boldsymbol{h}_i\in\bigcup_{z\in Z} \overline{\boldsymbol{H}}_{i,z}^*\}$. It can be shown that two informationally equivalent gradual mechanisms share the same instantized information flow for each agent $i$. Fourth, construct a unique immediate gradual mechanism $G^*$ from the instantized information flow $\{\mathbb{F}_i^*\}_{i\in N}$. See \ref{sec:Instantizing} for details of this construction. This establishes the first part of the theorem.

The proof of the second part of the theorem draws upon \citeauthor{Battigalli2020}'s \citeyearpar{Battigalli2020} characterization of behavioral equivalence. We first observe that the transformations of coalescing and simultanizing preserve informational equivalence. Therefore, behavioral equivalence between two gradual mechanisms implies they are also informationally equivalent. From their theorem, we know that each gradual mechanism is behaviorally equivalent to a unique gradual mechanism that has a ``minimal'' game structure---no opportunity for further coalescing and simultanizing. Notice that a gradual mechanism is immediate if and only if its game structure is minimal. It follows that a gradual mechanism $G$ must be behaviorally equivalent to the unique immediate gradual mechanism that is informationally equivalent with itself. This formally delivers an equivalence relation between informational equivalence and behavioral equivalence among gradual mechanisms and completes the proof. 

\end{proof}

To appreciate the importance of this theorem, notice that by combining the observations that have been individually made for behavioral equivalence and for gradual mechanisms in the context of dynamic implementation, the class of immediate gradual mechanisms could serve as a refined revelation principle under various solution concepts including Bayes-Nash, weak dominance, obvious dominance together with its generalizations using the sure-thing principle.

\section{Future Work}
We share in this paper some observations about behavioral equivalence in the setting of dynamic implementation. One can view such observations as guidance for the mechanism designer to navigate through local modifications on a dynamic game form. In this regard, it would be valuable to study other elementary transformations not necessarily maintaining informational equivalence while still implementing the social choice function. 

In this paper, the concept of information flow plays a central role in enabling the identification of a gradual mechanism with its unique informationally equivalent immediate gradual mechanism. This suggests an avenue for further research concerning the question of how to manage information flow in dynamic mechanisms. In this regard, issues involving privacy, confidentiality, credibility, and trust appear pertinent.

\newpage
\singlespacing
\bibliographystyle{chicago}
\bibliography{IDDM}

\begin{thebibliography}{}

\bibitem[\protect\citeauthoryear{Akbarpour and Li}{Akbarpour and
  Li}{2020}]{Akbarpour2020}
Akbarpour, M. and S.~Li (2020).
\newblock {Credible Auctions: A Trilemma}.
\newblock {\em Econometrica\/}~{\em 88}, 425--467.

\bibitem[\protect\citeauthoryear{Battigalli, Leonetti, and
  Maccheroni}{Battigalli et~al.}{2020}]{Battigalli2020}
Battigalli, P., P.~Leonetti, and F.~Maccheroni (2020).
\newblock {Behavioral Equivalence of Extensive Game Structures}.
\newblock {\em Games and Economic Behavior\/}~{\em 121}, 533--547.

\bibitem[\protect\citeauthoryear{B{\'{o}} and Hakimov}{B{\'{o}} and
  Hakimov}{2021}]{Bo2021}
B{\'{o}}, I. and R.~Hakimov (2021).
\newblock {Pick-an-object Mechanisms}.
\newblock Working Paper.

\bibitem[\protect\citeauthoryear{Bonannot}{Bonannot}{1992}]{Bonannot1992}
Bonannot, G. (1992).
\newblock {Set-Theoretic Equivalence of Extensive-Form Games}.
\newblock {\em International Journal of Game Theory\/}~{\em 20}, 429--447.

\bibitem[\protect\citeauthoryear{Chew and Wang}{Chew and Wang}{2022}]{CW2022}
Chew, S.~H. and W.~Wang (2022).
\newblock {Generalizing Obvious Dominance Using the Sure-thing Principle}.
\newblock Working Paper.

\bibitem[\protect\citeauthoryear{Elmes and Reny}{Elmes and
  Reny}{1994}]{Elmes1994}
Elmes, S. and P.~J. Reny (1994).
\newblock {On the Strategic Equivalence of Extensive Form Games}.
\newblock {\em Journal of Economic Theory\/}~{\em 62}, 1--23.

\bibitem[\protect\citeauthoryear{Harsanyi}{Harsanyi}{1967}]{Harsanyi1967}
Harsanyi, J.~C. (1967).
\newblock {Games with Incomplete Information Played by "Bayesian" Players,
  I-III: Part I. The Basic Model}.
\newblock {\em Management Science\/}~{\em 14}, 159--182.

\bibitem[\protect\citeauthoryear{Haupt and Hitzig}{Haupt and
  Hitzig}{2022}]{Haupt2022}
Haupt, A. and Z.~Hitzig (2022).
\newblock {Contextually Private Implementation}.
\newblock Working Paper.

\bibitem[\protect\citeauthoryear{Kohlberg and Mertens}{Kohlberg and
  Mertens}{1986}]{Kohlberg1986}
Kohlberg, E. and J.-F. Mertens (1986).
\newblock {On the Strategic Stability of Equilibria}.
\newblock {\em Econometrica\/}~{\em 54}, 1003--1037.

\bibitem[\protect\citeauthoryear{Kreps and Wilson}{Kreps and
  Wilson}{1982}]{Kreps1982}
Kreps, D.~M. and R.~Wilson (1982).
\newblock {Sequential Equilibria}.
\newblock {\em Econometrica\/}~{\em 50}, 863--894.

\bibitem[\protect\citeauthoryear{Kuhn}{Kuhn}{1950}]{Kuhn1950}
Kuhn, H.~W. (1950).
\newblock {Extensive Games}.
\newblock {\em Proceedings of the National Academy of Sciences of the United
  States of America\/}~{\em 36}, 570--576.

\bibitem[\protect\citeauthoryear{Li}{Li}{2017}]{Li2017}
Li, S. (2017).
\newblock {Obviously Strategy-Proof Mechanisms}.
\newblock {\em American Economic Review\/}~{\em 107}, 3257--3287.

\bibitem[\protect\citeauthoryear{Mackenzie}{Mackenzie}{2020}]{Mackenzie2020}
Mackenzie, A. (2020).
\newblock {A Revelation Principle for Obviously Strategy-proof Implementation}.
\newblock {\em Games and Economic Behavior\/}~{\em 124}, 512--533.

\bibitem[\protect\citeauthoryear{Mackenzie and Zhou}{Mackenzie and
  Zhou}{2022}]{MZ2022}
Mackenzie, A. and Y.~Zhou (2022).
\newblock {Menu Mechanisms}.
\newblock Working Paper.

\bibitem[\protect\citeauthoryear{Osborne and Rubinstein}{Osborne and
  Rubinstein}{1994}]{osborne1994course}
Osborne, M.~J. and A.~Rubinstein (1994).
\newblock {\em A Course in Game Theory}.
\newblock MIT Press.

\bibitem[\protect\citeauthoryear{Pycia and Troyan}{Pycia and
  Troyan}{2021}]{Pycia2021}
Pycia, M. and P.~Troyan (2021).
\newblock {A Theory of Simplicity in Games and Mechanism Design}.
\newblock Working Paper.

\bibitem[\protect\citeauthoryear{Selten}{Selten}{1975}]{Selten1975}
Selten, R. (1975).
\newblock {Reexamination of the Perfectness Concept for Equilibrium Points in
  Extensive Games}.
\newblock {\em International Journal of Game Theory\/}~{\em 4}, 25--55.

\bibitem[\protect\citeauthoryear{Thompson}{Thompson}{1952}]{Thompson1952}
Thompson, F.~B. (1952).
\newblock {Equivalence of Games in Extensive Forms}.
\newblock RAND Corporation Working Paper RM-759.

\end{thebibliography}

\newpage
\appendix
\renewcommand\thesection{Appendix \Alph{section}}
\onehalfspacing

\section{Definitions of Dynamic Game Form}
\label{sec:app_definitions}

\noindent In a dynamic game form $G = (\overline{H}, \{A_i, \boldsymbol{H}_i\}_{i\in N}, \mathcal{X})$ with perfect recall, we have:

\begin{itemize}
    \item \emph{Players.} Each agent $i\in N$ is a player.
    \item \emph{Actions.} For each agent $i\in N$, $A_i$ is a nonempty action set. Denote the set of action profiles by \[A = \{\varnothing\}\cup \bigcup_{\varnothing\subsetneq M\subseteq N} \prod_{i\in M} A_i\] in which the empty action profile $\varnothing$ is introduced only to simplify exposition.
    \begin{itemize}
        \item Pick an action profile $a\in \prod_{i\in M} A_i\subseteq A$. Suppose $i\in M$. Let $a_i$ denote the action of agent $i$ in $a$ and $a_{-i}$ the action profile of other agents in $a$. Otherwise, suppose $i\notin M$. Then $a_i=\varnothing$ and $a_{-i} = a$.
        \item For each $T > 0$, let $A^T$ denote the collection of histories of length $T$ with a generic history being denoted by $h = (h^{(1)}, \ldots, h^{(T)})$ in which $h^{(T)}$ is also referred to as $h^{(-1)}$. Let $A^0 = \{\varnothing\}$ be the singleton set of the empty history.
        \item Let $A^{<\mathbb{N}} = \bigcup_{T\in \mathbb{N}} A^T$ denote the collection of all histories of finite length.
        \item An action profile $a$ and the corresponding sequence $(a)$ containing only $a$ are used interchangeably. The empty history $\varnothing$ and any sequence of empty action profiles are used interchangeably. For history $h = (h^{(1)}, \ldots, h^{(T)})$ consisting of $S$ non-empty action profiles and some empty action profiles, let $\underline{h} = (\underline{h}^{(1)}, \ldots, \underline{h}^{(S)})$ be the corresponding history without empty action profiles, i.e., $s$-th non-empty action profiles in $h$ equals $\underline{h}^{(s)}$ for each $1 \leq s \leq S$. We will use $h$ and $\underline{h}$ interchangeably.
        \item There is a precedence relation $\preceq$ on $A^{<\mathbb{N}}$, i.e., $\uline{h}\preceq h$ (reads $\uline{h}$ is a predecessor of $h$ or $h$ is a successor of $\uline{h}$) if $\uline{h}\in A^S$ and $h\in A^T$ such that $S = 0$ or that $0 < S \leq  T$ and $\uline{h}^{(s)} = h^{(s)}$ for any $1\leq s\leq S$.
        \item Let $h_1,\ldots, h_m\in A^{<\mathbb{N}}$ be $m$ sequences of action profiles, define $(h_1, \ldots, h_m)\in A^{<\mathbb{N}}$ by concatenation.
        \item If $\uline{h}\preceq h$ in $A^{<\mathbb{N}}$ such that $\uline{h}\in A^T$, $h\in A^{T + 1}$, and $h^{(-1)}\neq\varnothing$, we say $\uline{h}$ is an immediate predecessor of $h$ or $h$ is an immediate successor of $\uline{h}$. Note that a nonempty history $h$ has a unique immediate predecessor.
        \item Let $h = (h^{(1)}, \ldots, h^{(T)})$, define $h_{i} = (h^{(1)}_i, \ldots, h^{(T)}_i)$ and $h_{-i} = (h^{(1)}_{-i}, \ldots, h^{(T)}_{-i})$.
    \end{itemize}
    \item \emph{Histories.} The set of histories $\overline{H}$ is  modeled by a tree of finite length in $A^{< \mathbb{N}}$, i.e., a subset of $A^{< \mathbb{N}}$ such that (i) there exists $\mathcal{T}$ such that $\overline{H}\subseteq \bigcup_{T=0}^\mathcal{T} A^T$, (ii) $\varnothing\in \overline{H}$, and (iii) for any $h\in \overline{H}$ such that $h\neq \varnothing$, the immediate predecessor of $h$ is in $\overline{H}$. We make use of the following assumptions and notations.
    \begin{itemize}
        \item Denote the set of terminal histories by $Z$ and non-terminal histories by $H$.
        \item For any non-terminal history $h\in H$, denote by $\sigma(h)$ the collection of its immediate successors.
        \item $\overline{H}$ satisfies the following property: there exists an active-player correspondence $\mathbb{P}: H\twoheadrightarrow N$ such that for any non-terminal history $h\in H$ and any $a\in A$ satisfying $(h, a)\in \sigma(h)$, it is the case that $a\in \prod_{i\in\mathbb{P}(h)} A_i$. Therefore, $\mathbb{P}(\cdot)$ depicts the players that are simultaneously active at a particular non-terminal history.
        \item Let $H_i = \{h\in H: i \in \mathbb{P}(h)\}$ represent the collection of histories on which player $i$ is active. For each $h\in H$ and each $i\in \mathbb{P}(h)$, define $A_i(h) = \{a_i\in A_i: (h, a)\in \overline{H} \mbox{ for some } a\in A\}$ as the collection of available actions for player $i$ at history $h$.
        \item $\overline{H}$ satisfies the following property: for any $h\in H$ and any $a\in \prod_{i\in \mathbb{P}(h)} A_i(h)$, we have $(h, a)\in \overline{H}$. The two assumptions here depict decentralized decision making.
    \end{itemize}
    \item \emph{Information Structure.} For each $i\in N$, $\boldsymbol{H}_i$ is a partition of $H_i$ whose elements are information sets of player $i$, with a generic information set being denoted by $\boldsymbol{h}_i$. We introduce further assumptions, notations, and observations below.
    \begin{itemize}
        \item Assume that for any $\boldsymbol{h}_i\in \boldsymbol{H}_i$ and any $h, \tilde{h}\in \boldsymbol{h}_i$, we have $A_i(h) = A_i(\tilde{h})$. Then $A(\boldsymbol{h}_i) = A_i(h)$ for which $h\in \boldsymbol{h}_i$ is well defined.
        \item Assume that the game form $G$ has perfect recall. Formally, for any $h, \tilde{h}\in \boldsymbol{h}_i$, we have $h_i = \tilde{h}_i$ and for any $\underline{h}\preceq h$ with $\underline{h}\in H_i$, there exists $\underline{\tilde{h}}\preceq \tilde{h}$ such that $\underline{h}$ and $\underline{\tilde{h}}$ are in the same information set of agent $i$.
        
        \item Given perfect recall, an ordering on $\boldsymbol{H}_i$ can be defined, $\uline{\boldsymbol{h}}_i\preceq \boldsymbol{h}_i$, if there exist $\uline{h}\in \uline{\boldsymbol{h}}_i$ and $h\in \boldsymbol{h}_i$ such that $\uline{h}\preceq h$.
    \end{itemize}
    \item \emph{Outcomes.} $\mathcal{X}:Z\rightarrow X$ assigns each terminal history a public outcome.
\end{itemize}


For agent $i\in N$, an interim strategy $s_i: H_i\rightarrow A_i$ specifies an available action $s_i(h)\in A_i(h)$ for each history $h\in H_i$ such that $s_i(h) = s_i(\tilde{h})$ if $h, \tilde{h}$ belong to the same information set. Therefore, $s_i: \boldsymbol{H}_i\rightarrow A_i$ is well defined. We use $S_i$ to denote the set of interim strategies for agent $i$ and use $S$ and $S_{-i}$ to denote respectively the profile of interim strategies for all agents in $N$ and for those other than agent $i$.

\section{Coalescing and Simultanizing}
\label{sec:Elementary_Transformation}

\noindent We describe here our adaptation of the transformations of coalescing and simultanizing from  \cite{Battigalli2020} on gradual mechanisms.

\subsection{Coalescing}

Intuitively, the coalescing transformation applies to two adjacent information sets $\underline{\boldsymbol{h}}_i^c$ and $\boldsymbol{h}_i^c$ of agent $i$. This transformation advances the available actions $A_i(\boldsymbol{h}_i^c)$ of agent $i$ on the latter information set $\boldsymbol{h}_i^c$ to the prior one, replacing the action $\Theta_i(\boldsymbol{h}_i^c)$ at information set $\underline{\boldsymbol{h}}_i^c$ leading to the latter information set. This can be done in a way without changing any agents' information flow if agent $i$ does not learn any new information upon arriving at the latter information set $\boldsymbol{h}_i^c$.

Formally, there is an opportunity for coalescing in gradual mechanism $G$ if there exists an agent $i$, two information sets $\boldsymbol{h}_i^c$ and $\underline{\boldsymbol{h}}_i^c$ in $\boldsymbol{H}_i$, and $\underline{a}_i\in A_i(\underline{\boldsymbol{h}}_i^c)$ such that $\sigma_{\underline{a}_i}(\underline{\boldsymbol{h}}_i^c) = \{\boldsymbol{h}_i^c\}$, or equivalently, $\boldsymbol{h}_i^c\in \sigma(\underline{\boldsymbol{h}}_i^c)$, $\Theta_i(\boldsymbol{h}_i^c) = \underline{a}_i$, and $\Theta_{-i}(\boldsymbol{h}_i^c) = \Theta_{-i}(\underline{\boldsymbol{h}}_i^c)$. When there exists such an opportunity in $G$, coalescing delivers $G^*$ in the following manner.

\begin{itemize}
    \item Each history $h^*\in\overline{H}^*$ corresponds to $m(h^*)\in\overline{H}$ in one of two ways:\footnote{A constructive definition of $\overline{H}^*$ starting from the initial history can be given as well. Instead, we describe the mapping $m:\overline{H}^*\rightarrow \overline{H}$ as an equivalent characterization of $\overline{H}^*$. This mapping can also be used in describing information sets and outcome function of $G^*$. We apply the same method to define the transformation of simultanizing.}
    \begin{enumerate}
        \item $m(h^*) = h^*$ if it is not the case that $(h, (a_i, a_{-i}))\preceq h^*$ for some $h\in \underline{\boldsymbol{h}}_i^c$, $a_i\in A_i(\boldsymbol{h}_i^c)$, and $a_{-i}\in A$.
        \item $m(h^*) = (h, (\underline{a}_i, a_{-i}), h', h'')$ if $h^* = (h, (a_i, a_{-i}), h'_{-i}, h'')$ for some $h\in\underline{\boldsymbol{h}}_i^c$, $a_i\in A_i(\boldsymbol{h}_i^c)$, $a_{-i}\in A$, and $h', h''\in A^{<\mathbb{N}}$ with either (i) $h'_i = h''_i = \varnothing$ or (ii) $h'_{i} = a_i$.
    \end{enumerate}
    \item For each agent $j\in N$, the information sets of agent $j$ in $G^*$ is given by the non-empty sets in $\{\{h^*\in H_j^*: m(h^*) \in \boldsymbol{h}_j\}\}_{\boldsymbol{h}_j\in\boldsymbol{H}_j}$. Notice that only $\{h^*\in H_i^*: m(h^*) \in \boldsymbol{h}_i^c\}$ is an empty set, i.e., only the information set corresponding to $\boldsymbol{h}_i^c$ is eliminated from $G^*$.
    \item The outcome function $\mathcal{X}^*$ is given by $\mathcal{X}^*(z^*) = \mathcal{X}(m(z^*))$.
\end{itemize}

\subsection{Simultanizing}

Like the coalescing transformation, simultanizing advances decisions to an ealier point $\underline{h}$ from a later stage $\boldsymbol{h}_i^s$. Suppose the administrator has accrued enough information to compose a new message for agent $i$ at $\underline{h}$, then she could contact agent $i$ immediatly rather than waiting to do so at a subsequent history in $\boldsymbol{h}_i^s$.

Formally, there is an opportunity for simultanizing if there exist $\underline{h}\in H$ and $\boldsymbol{h}_i^s\in\boldsymbol{H}_i$ such that (i) $\underline{h}\notin H_i$, (ii) $\underline{h}\prec h$ for some $h\in\boldsymbol{h}_i^s$ with $\underline{h}_i = h_i$, and (iii) $\Theta_{-i}(h)\subseteq \Theta_{-i}(\boldsymbol{h}_i^s)$. When such an opportunity exists, simultanizing on $G$ yields $G^*$ in the following way.

\begin{itemize}
    \item Each history $h^*\in\overline{H}^*$ corresponds to a history $m(h^*)\in\overline{H}$ in one of two ways:
    \begin{enumerate}
        \item $m(h^*) = h^*$ if it is not the case that $h\prec h^*$.
        \item $m(h^*) = (h, a_{-i}, h', h'')$ if $h^* = (h, (a_i, a_{-i}), h'_{-i}, h'')$ for some $a_i\in A_i(\boldsymbol{h}_i^s)$, $a_{-i}\in A$, and $h', h''\in A^{<\mathbb{N}}$ with either (i) $h'_i = h''_i = \varnothing$ or (ii) $h'_{i} = a_i$.
    \end{enumerate}
    \item For each agent $j\in N$, the information sets of agent $j$ in $G^*$ are given by $\{\{h^*\in H_j^*: m(h^*) \in \boldsymbol{h}_j\}\}_{\boldsymbol{h}_j\in\boldsymbol{H}_j}$.
    \item The outcome function $\mathcal{X}^*$ is given by $\mathcal{X}^*(z^*) = \mathcal{X}(m(z^*))$.
\end{itemize}

\section{The Fourth Step of Instantizing}
\label{sec:Instantizing}

For agent $i$, each element in the instantized information flow $\mathbb{F}_i^*$ can be written as a product $P_i\times R_{-i}$ of the provided information $P_i\subseteq \Theta_i$ and the recieved information $R_{-i}\subseteq \Theta_{-i}$. By Proposition \ref{prop:Flows} and the first three steps, each of the original $\mathbb{F}_i^*$ forms an arborescence under the binary relation $\supseteq$ with the collection of immediate successors of each $P_i\times R_{-i}$ forming a partition of $P_i\times R_{-i}$ and the sub-collection sharing the same provided information $P_i'\subseteq P_i$ forming a partition of $P_i'\times R_{-i}$. Let $\mathbb{G}_i^*$ be the subset of $\mathbb{F}_i^*$ consisting of its initial nodes with immediate successors. Removing from $\mathbb{F}_i^*$ some initial nodes in $\mathbb{G}_i^*$ defines a new arborescence with the same property.


We provide an iterative construction of the histories $\overline{H}^*$ of $G^*$ based on the instantized information flows $\{\mathbb{F}_i^*\}_{i\in N}$ using repeated removal of initial nodes in $\mathbb{F}_i^*$ of each agent and repeated addition of histories into $\overline{H}^*$. We begin our construction with the original $\{\mathbb{F}^*_i\}_{i\in N}$ at which $\overline{H}^* = \{\varnothing\}$. At each step, let $h$ be a terminal history of $\overline{H}^*$ under construction, in which case $\mathbb{P}(h) = \{i\in N: \exists P_i\times R_{-i}\in\mathbb{G}_i^* \mbox{, } \Theta(h)\subseteq P_i\times R_{-i}\}$ is not empty. Notice that for each agent $i\in \mathbb{P}(h)$, there exists a unique $P_i\times R_{-i}\in \mathbb{G}_i^*$ such that $\Theta(h)\subseteq P_i\times R_{-i}$.

\begin{itemize}
    \item For each $i\in \mathbb{P}(h)$, let $P_i\times R_{-i}$ be the unique initial node containing $\Theta(h)$. Define $A_i(h)$ to be the collection of $a_i$ such that $a_i\times R_{-i}'$ is an immediate successor of $P_i\times R_{-i}$ in $\mathbb{F}_i^*$.
    \item For each $(a_i)_{i\in \mathbb{P}(h)}$ in which $a_i\in A_i(h)$, add $(h, (a_i)_{i\in \mathbb{P}(h)})$ to $\overline{H}^*$.
    \item For each $i\in \mathbb{P}(h)$, remove the unique $P_i\times R_{-i}$ such that $\Theta(h)\subseteq P_i\times R_{-i}$ from $\mathbb{F}_i^*$.
\end{itemize}

Continue this process until $\mathbb{P}(h)$ is empty for each terminal history of $\overline{H}^*$. Define information sets of $G^*$ such that information flows in $G^*$ are the original instantized information flows $\mathbb{F}_i^*$. Define outcome function of $G^*$ such that $\mathcal{X}^*(z^*) = \mathcal{X}(z)$ if $\Theta(z^*) = \Theta(z)$.

\end{document}